\pdfoutput=1 

\documentclass[10pt]{article}
\usepackage{amssymb}

\usepackage{amsthm} %for \theoremstyle{definition}

\usepackage{tikz}
\usetikzlibrary{trees}
\usepackage{qtree}

\usetikzlibrary{decorations.pathmorphing}
\usetikzlibrary{decorations.markings}
\usepackage{verbatim}

\usepackage{amsmath}
\usepackage{amsthm} %for \theoremstyle{definition}
\usepackage{amssymb,amsfonts,textcomp}
\usepackage{array}
\usepackage{hhline}
\usepackage{graphicx}

\usepackage[T3,T1]{fontenc} %for \textquotedbl
\usepackage{txfonts} %/pxfonts for \circleright

\usepackage{latexsym} %for \Box
\usepackage{mathrsfs} %for \mathscr

\usepackage{MnSymbol} %for more curved arrow symbol and many others are possible \lcurvearrowright

\usepackage{tikz}
\usetikzlibrary{arrows,shapes,chains,matrix,positioning,scopes}% shapes for circle split
\usetikzlibrary{decorations.pathmorphing}
\usetikzlibrary{decorations.markings}
\usepgflibrary{plotmarks}%for squares and other shapes on graphs lines , 
\usetikzlibrary{patterns}
\usepackage{pdfpages}

\usetikzlibrary{mindmap}%for circle connection bar

\usepackage{graphicx}

\usepackage{multirow}
\usepackage{multicol}

\usepackage{subfig} 

\usepackage{float}
\usepackage{wrapfig}
\restylefloat{figure}
\restylefloat{table}
\usepackage{color, colortbl} %to color rows
\usepackage{float}
\restylefloat{figure}

\theoremstyle{definition}

\definecolor{currentcolor}{rgb}{0.8 0.4 0.2}%orange currentcolor is to be set to path color and then made lighter

\tikzstyle{stochasticjumpstyle}=[diamond,draw,fill=white,>=latex,>->,dashed]
\tikzstyle{stochasticPathstyle}=[>=latex,>->,dashed]
\tikzstyle{stochasticNodestyle}=[ellipse,inner sep=1pt,text=.,fill=.!20]%[fill=white,inner sep=1pt]%[ellipse,inner sep=1pt,draw,fill=white]
\tikzstyle{blankstyle}=[fill=white,inner sep=1pt]

\def\SnakeSegLen{0.6em}%defines snake segment length for signal jumps  in graphs
\def\SnakeAmp{0.11em}%defines snake amplitude for signal jumps  in graphs
\def\PrePostLen{5mm}

\tikzstyle{sendstyle}=[dashed,line width=1.1pt]%[dotted,ultra thick]

\tikzstyle{splitstyle}=[circle,draw]%not used

\tikzstyle{receivestyle}=[>->,line width=1.1pt,decorate, decoration={zigzag,segment length=\SnakeSegLen, amplitude=\SnakeAmp, pre length=\PrePostLen, post=curveto, post length=\PrePostLen},text=black]

\tikzstyle{receivesigstyle}=[draw,inner sep=2pt,fill=pink!20]
\tikzstyle{receivesigstyle3}=[draw,inner sep=2pt, fill=white]
\tikzstyle{receivesigstyle2}=[ellipse,shade, draw,double,fill=red!10]

\tikzstyle{sendsigstyle}=[diamond,draw,inner sep=1pt, text=black, fill=yellow!80]
\tikzstyle{sendsigstyle3}=[circle,draw, ball color=white]
\tikzstyle{sendsigstyle2}=[diamond,draw,double, inner sep=1pt, fill=white]

\tikzstyle{snakesendstyle}=[*->, decorate, decoration={snake, segment length=\SnakeSegLen, amplitude=\SnakeAmp,  pre length=\PrePostLen, post=curveto, post length=\PrePostLen}]

\tikzstyle{snakesendstyle1}=[line width=1.1pt, decorate, decoration={snake,segment length=\SnakeSegLen, amplitude=\SnakeAmp}]

\tikzstyle{snakesendstyle3}=[decorate, decoration={markings, mark=at position .75 with {\arrow[red,line width=5mm]{>}}, snake, segment length=\SnakeSegLen, amplitude=\SnakeAmp,  pre length=\PrePostLen, post=curveto, post length=\PrePostLen}]

\tikzstyle{snakesendstyle2}=[decorate, decoration={ zigzag,segment length=\SnakeSegLen, amplitude=\SnakeAmp, line around/.style={decoration={pre length=\PrePostLen,post length=\PrePostLen}}}]
\newcounter{foo}
\usepackage[bookmarks=false]{hyperref} %To prevent error in arXiv 

\makeatletter
\AtBeginDocument{\let\nameref\HyPsd@nameref}

\makeatother

\colorlet{anglecolor}{green!50!black}
\definecolor{darkgreen}{rgb}{0 0.6  0}

\definecolor{turquoise}{rgb}{0 0.41 0.41}
\definecolor{rouge}{rgb}{0.79 0.0 0.1}
\definecolor{vert}{rgb}{0.15 0.4 0.1}
\definecolor{mauve}{rgb}{0.6 0.4 0.8}
\definecolor{violet}{rgb}{0.58 0. 0.41}
\definecolor{orange}{rgb}{0.8 0.4 0.2}
\definecolor{bleu}{rgb}{0.39, 0.58, 0.93}

\definecolor{darkross}{rgb}{0.008,0.412,0.471}
\definecolor{middleross}{rgb}{0.012,0.580,0.663}
\definecolor{lightross}{rgb}{0.016,0.749,0.855}
\definecolor{darkblue}{rgb}{0.067,0.008,0.471}
\definecolor{middleblue}{rgb}{0.094,0.012,0.663}
\definecolor{lightblue}{rgb}{0.122,0.016,0.855}
\definecolor{darkpurple}{rgb}{0.471,0.008,0.412}
\definecolor{middlepurple}{rgb}{0.663,0.012,0.580}
\definecolor{lightpurple}{rgb}{0.855,0.016,0.749}
\definecolor{darkbrown}{rgb}{0.471,0.067,0.008}
\definecolor{middlebrown}{rgb}{0.663,0.094,0.012}
\definecolor{lightbrown}{rgb}{0.855,0.122,0.016}
\definecolor{darkolive}{rgb}{0.412,0.471,0.008}
\definecolor{middleolive}{rgb}{0.580,0.663,0.012}
\definecolor{lightolive}{rgb}{0.749,0.855,0.016}
\definecolor{darkgreen}{rgb}{0.008,0.417,0.067}
\definecolor{middlegreen}{rgb}{0.012,0.663,0.094}
\definecolor{lightgreen}{rgb}{0.016,0.855,0.122}
\definecolor{darkocre}{rgb}{0.471,0.298,0.008}
\definecolor{middleocre}{rgb}{0.663,0.420,0.012}
\definecolor{lightocre}{rgb}{0.855,0.541,0.016}

    \definecolor{lightblue}{rgb}{0,0,.7}
    \definecolor{orange}{rgb}{1,.7,0}
    \definecolor{darkorange}{rgb}{1,.4,0}
    \definecolor{darkgreen}{rgb}{0,.5,0}
    \definecolor{darkblue}{rgb}{0,0,.4}
    \definecolor{darkred}{rgb}{.4,0,0}
    \definecolor{gray}{rgb}{.2,.2,.2}
    \definecolor{darkgray}{rgb}{.2,.2,.2}
    \definecolor{shadecolor}{gray}{0.925}
    
\definecolor{darkred}{rgb}{0.65,0,0}
\definecolor{darkblue}{rgb}{0,0,.65}
\definecolor{darkgreen}{rgb}{0,0.5,0}
\definecolor{orange}{rgb}{1,.75,.25}
\definecolor{aqua}{rgb}{0,.25,.75}
\definecolor{grey}{rgb}{.5,.5,.5}
\definecolor{brown}{rgb}{.51,.35,.18}

\definecolor{lightblue}{rgb}{.3,.5,1}
\definecolor{orange}{rgb}{1,.7,0}
\definecolor{darkorange}{rgb}{1,.4,0}
\definecolor{darkgreen}{rgb}{0,.4,0}
\definecolor{darkblue}{rgb}{0,0,.4}
\definecolor{darkred}{rgb}{.56,0,0}
\definecolor{gray}{rgb}{.3,.3,.3}
\definecolor{darkgray}{rgb}{.2,.2,.2}

\definecolor{blue}{rgb}{0,0,1}
\definecolor{red}{rgb}{1,0,0}
\definecolor{pink}{rgb}{.933,0,.933}
\definecolor{green}{rgb}{0.133,0.545,0.133}
\definecolor{shadecolor}{gray}{0.925}

\definecolor{DarkBlue}{rgb}{0.000,0.000,0.545}
\definecolor{DarkChocolate}{rgb}{0.400,0.200,0.000}
\definecolor{DarkCyan}{rgb}{0.000,0.545,0.545}
\definecolor{DarkGoldenrod}{rgb}{0.720,0.525,0.044}
\definecolor{DarkGray}{rgb}{0.664,0.664,0.664}
\definecolor{DarkGreen}{rgb}{0.000,0.392,0.000}
\definecolor{DarkGrey}{rgb}{0.664,0.664,0.664}
\definecolor{DarkKhaki}{rgb}{0.740,0.716,0.420}
\definecolor{DarkLavender}{rgb}{0.400,0.200,0.600}
\definecolor{DarkMagenta}{rgb}{0.545,0.000,0.545}
\definecolor{DarkOliveGreen}{rgb}{0.332,0.420,0.185}
\definecolor{DarkOrange}{rgb}{1.000,0.550,0.000}
\definecolor{DarkOrchid}{rgb}{0.600,0.196,0.800}
\definecolor{DarkPeriwinkle}{rgb}{0.400,0.400,1.000}
\definecolor{DarkPurpleBlue}{rgb}{0.400,0.000,0.800}
\definecolor{DarkRed}{rgb}{0.545,0.000,0.000}
\definecolor{DarkRoyalBlue}{rgb}{0.000,0.200,0.800}
\definecolor{DarkSalmon}{rgb}{0.912,0.590,0.480}
\definecolor{DarkSeaGreen}{rgb}{0.560,0.736,0.560}
\definecolor{DarkSlateBlue}{rgb}{0.284,0.240,0.545}
\definecolor{DarkSlateGray}{rgb}{0.185,0.310,0.310}
\definecolor{DarkSlateGrey}{rgb}{0.185,0.310,0.310}
\definecolor{DarkSmoke}{rgb}{0.920,0.920,0.920}
\definecolor{DarkTurquoise}{rgb}{0.000,0.808,0.820}
\definecolor{DarkViolet}{rgb}{0.580,0.000,0.828}
\definecolor{DeepPink}{rgb}{1.000,0.080,0.576}
\definecolor{DeepSkyBlue}{rgb}{0.000,0.750,1.000}

\tikzstyle{mystyle}=[scale= \PicSize,  %[****Crit. PicSize is not defined*****]
baseline=(current bounding box.center),
nodedescr/.style={fill=white,inner sep=2.5pt},
nodedescr2/.style={draw,circle,fill=white},
description/.style={fill=white,inner sep=2pt},
cancer loop/.style={preaction={draw=white, -,line width=6pt}, line width=1.1pt},
pot1 loop/.style={preaction={draw=white, -,line width=6pt}, red, line width=1.1pt},
pot2 loop/.style={preaction={draw=white, -,line width=6pt}, green, line width=1.1pt},
pot1/.style={preaction={draw=white, -,line width=6pt}, gray, solid, line width=1pt}, 
pot2/.style={preaction={draw=white, -,line width=6pt}, gray, solid, line width=1pt},
inPot/.style={ out=45,in=90,distance=3cm, min distance=2cm, line width=1.1pt, black!75},
inPot1/.style={ out=45,in=120,distance=3cm, min distance=2cm, line width=1.1pt, black!75},
inPot2/.style={ out=-45,in=-120,distance=3cm, min distance=2cm, line width=1.1pt, black!75},
in100Pot1/.style={ out=80,in=90,distance=3cm, min distance=2cm, line width=1.1pt, darkgreen},
in100Pot2/.style={ out=-80,in=-90,distance=3cm, min distance=2cm, line width=1.1pt, brown},
in2Pot1/.style={ out=45,in=90,distance=3cm, min distance=2cm, line width=1.1pt, black!75},
in2Pot2/.style={ out=-65,in=-90,distance=3cm, min distance=2cm, line width=1.1pt, black!75},
selfloop1/.style={ out=45,in=120,distance=6cm, min distance=4cm, line width=1.1pt, red},
selfloop2/.style={ out=-45,in=-120,distance=6cm, min distance=4cm, line width=1.1pt, blue},
loop1red/.style={ out=45,in=120,distance=6cm, min distance=4cm, line width=1.1pt, red},
loop2red/.style={ out=-45,in=-80,distance=6cm, min distance=4cm, line width=1.1pt, red},
cross line/.style={preaction={draw=white, -,	line width=6pt}},
jump1/.style={ out=45,in=135,distance=4cm, min distance=2cm, line width=1.1pt, red, dashed},
jump2/.style={ out=-45,in=-135,distance=4cm, min distance=2cm, line width=1.1pt, blue, dashed},
jumpover/.style={ out=80,in=90,distance=4cm, min distance=3cm, line width=1.1pt, black, dashed},
stochastic jump/.style={>=latex,>->,dashed, line width=1.1pt, green}
]

\def\PicSzThree{0.3}

\def\PicSzGyn{0.27}

\def\PicSize{0.5} % 0.5 defines constant PicSize for uniform scale of TikZ pictures

\def\nexttoPicSize2{6.0cm}

\def\oriPicSize{3.3cm}

\numberwithin{equation}{section}
\usepackage[margin=1.4in]{geometry} %***set page dimensions easily. See geometryPageSize.pdf***

\usepackage[parfill]{parskip}    % Activate to begin paragraphs with an empty line rather than an indent

\begin{document}

\title{A Developmental Network Theory of Gynandromorphs, Sexual Dimorphism and Species Formation}
% using Developmental Control Network Theory}
%\title{Computational Modeling of Developmental Control Networks for Gynandromorphs, Sexual Dimorphism and Species Formation}
%: \\ A Computational Developmental Control Network Theory}

\author{Eric Werner \thanks{Balliol Graduate Centre, Oxford Advanced Research Foundation (http://oarf.org), Cellnomica, Inc. (http://cellnomica.com). Thanks: Martin Brasier, Mike Clinton, Richard Gardner, Donald Hall, Francis Hitching, Rich Palmer, for images and/or helpful discussions.  
\copyright Eric Werner 2012.  All rights reserved. }\\ \\
University of Oxford\\
Department of Physiology, Anatomy and Genetics, \\
and Oxford Advanced Research Foundation, \\
Le Gros Clark Building, 
South Parks Road, 
Oxford OX1 3QX  \\
email:  eric.werner@dpag.ox.ac.uk\\
Website: http://ericwerner.com
}

\date{ } %This is to suppress the printing out of the date.

\maketitle

\thispagestyle{empty}

\begin{center}
\textbf{Abstract}

\begin{quote}
\it 
 Gynandromorphs are creatures where at least two different body sections are a different sex.  Bilateral gynandromorphs are half male and half female.  Here we develop a theory of gynandromorph ontogeny  based on developmental control networks.  The theory explains the embryogenesis of all known variations of gynandromorphs found in multicellular organisms.  The theory also predicts a large variety of more subtle gynandromorphic morphologies yet to be discovered. The network theory of gynandromorph development has direct relevance to understanding sexual dimorphism (differences in morphology between male and female organisms of the same species) and medical pathologies such as hemihyperplasia (asymmetric development of normally symmetric body parts in a unisexual individual). The network theory of gynandromorphs brings up fundamental open questions about developmental control in ontogeny. This in turn suggests a new theory of the origin and evolution of species that is based on cooperative interactions and conflicts between developmental control networks in the haploid genomes and epigenomes of potential sexual partners for reproduction.  This network-based theory of the origin of species is a paradigmatic shift in our understanding of evolutionary processes that goes beyond gene-centered theories. 
\end{quote}
\end{center}

\pagebreak

%\pagenumbering{roman}
%\setcounter{page}{1}
\tableofcontents
%%\listoffigures
%%\listoftables
%\newpage
%\pagenumbering{arabic}

\pagebreak

\section{Introduction}
\label{sec:Intro}

\begin{figure}[H]
\centering
\subfloat[Part 4][Lobster ]{\includegraphics[scale=0.108]{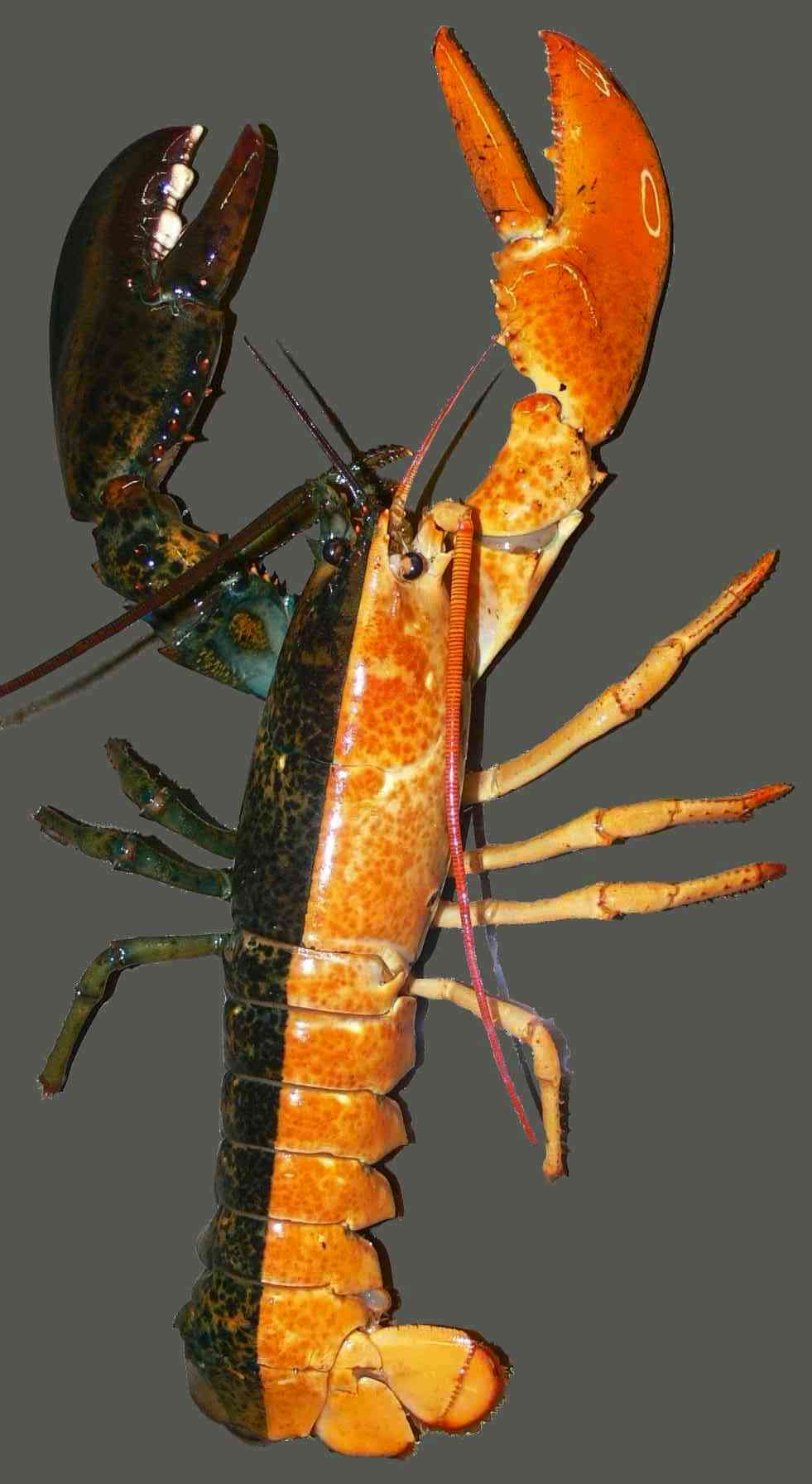} \label{fig:Lobster}}\hspace{1cm}
\subfloat[Part 1][Polar-O Mosquitoes]{\includegraphics[scale=0.35]{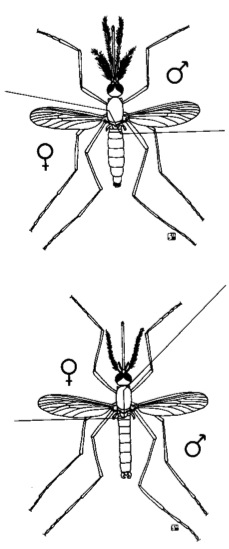} \label{fig:PolarObliqueMosquitoes}} \hspace{1cm}
\subfloat[Part 1][Rooster-Hen]{\includegraphics[scale=0.3]{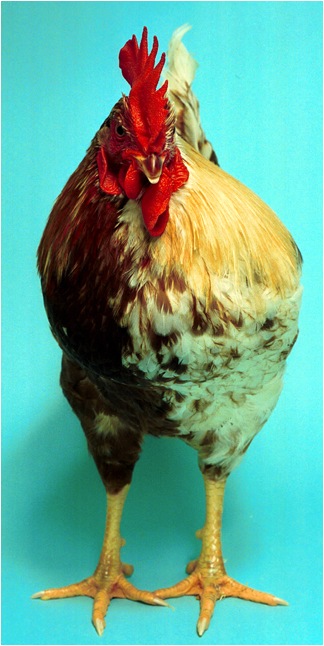} \label{fig:RoosterHen}} \hspace{1cm}
\caption{{\bf  Gynandromorphs.} \it \small 
Fig.\ref{fig:Lobster} Lobster, yellow male on the right hand side, green-brown female on the left hand side.  Picture provided by Palmer \cite{Palmer2009}. 
Fig.\ref{fig:PolarObliqueMosquitoes} Polar-Oblique mosquito illustrations. Used with permission from Hall \cite{Hall1988}. 
Fig.\ref{fig:RoosterHen} Rooster-Hen, its right side is female brown, left side male yellow. Picture provided by Clinton. A version appears in \cite{Zhao2010}.
 }
\label{fig:GynanderPics}
\end{figure}

Gynandromorphs are creatures that are part female (gyn-) and part male (andro-).  Bilateral gynandromorphs are half male and half female split down the middle like the lobster in Fig.\ref{fig:Lobster} and the Rooster-Hen in Fig.\ref{fig:RoosterHen}.  Gynandromorphs occur in many species of  including insects, fish, birds and mammals.  

Scientists give conflicting answers to the questions of how gynandromorphs form and what makes them possible at all.  The standard view used is that gynandromorphs result from an unequal distribution of sex chromosomes by way of chromosome loss or gain during cell division.  Others have found the cells to have a normal complement of chromosomes, but the cells in the two bilateral halves are of different sex. Related to this is the view is that hormones determine sexual dimorphism -the differences in morphology between males and females.  This paper gives an explanation of why chromosomal abnormalities and cells of different sex can lead to gynandromorphism.  In addition it is shown that there is  a third way for gynandromorphs to develop where all the cells of an organism are genetically normal with the normal complement of chromosomes and all the cells have the same sex, and yet they can still develop into gynandromorphs or have hemimorphic, non-bilateral bodies, 

\section{Networks control gynandromorph development}

This essay presents a novel, general computational theory of how gynandromorphs are generated from a single fertilized egg.  The theory is based on a computational theory of development, developmental control networks (CENEs) and bilateral symmetry \cite{Werner2003, Werner2005, Werner2011a, Werner2012a}.   The standard hypotheses that an unequal chromosome distribution causes gynandromorphy are special cases of the general network theory of gynandromorphic development presented here.   

What is missing in all previous accounts of gynandromorphs is a detailed explanation and theory of how gynandromorphs develop from a single fertilized egg. 
This paper gives an explanation of the ontogeny of gynandromorphs that is detailed enough to model the development of all known types of gynandromorphs found in multicellular organisms\footnote{When a theory cannot explain a phenomena it is simply ignored by the proponents of that theory.  Research on gynandromorphs was popular in the 1930's and later (see articles in the Journal of Heredity 1929-present). Then, with the change of biology's scientific paradigm to molecular biology, since gynandromorphs could not be explained genetically, the area was largely ignored.  However, with the developmental control network theory \cite{Werner2011a} together with the theory of bilateral symmetry \cite{Werner2012a}, gynandromorphs can now be understood. }.  The theory also predicts a large variety of more subtle gynandromorphic morphologies yet to be discovered.   

First I will give a brief introduction to developmental control networks (CENEs). Then I present some examples of types of gynandromorphs and relate them to the developmental control networks that generate them. It will be shown that each gynandromorph network has a corresponding meta-network signature that can be used to distinguish and classify gynandromorphs.  The study of gynandromorph developmental networks gives fascinating new insights into how species originate, which I will discuss at the end.
 
\section{Modeling gynandromorphs}
\label{ModelingGynanders}
One of the best ways of understanding the complex processes involved in the development of multicellular organisms is to computationally model the control network architecture of genomes and their interactions with cells. Cells interpret the developmental control networks in their genome and give that genome pragmatic meaning. The cell's essential internal orientation and its external orientation in space also needs to be modeled.  In addition, one also has to model the cell-cell communication and cell physics.  One can then run simulations of multicellular development in space and time and observe the resulting organism. The modeling also permits making changes to the network, running the simulation again and seeing the new resulting form.  Below I will describe the results of modeling gynandromorphs and the developmental control networks (CENEs) that lead to their development from starting from a single cell.  This will give us a deep insight into the nature of not only gynandromorphs but into the developmental control and evolution of all multicellular life. 

\subsection{Developmental Control Networks or CENEs}
To understand the dynamics of development I take a perspective that goes beyond the current gene-centered paradigm. My view is that the complexity of organisms and the related complexity of the ontogeny of embryos requires control information that is not in genes. Genes make up the interacting parts of the cell but they do not contain the information that orders, organizes and structures the dynamic development of multicellular systems.  Genes produce parts and processes that are local to the cell. {\em CENEs} (developmental control networks) contain the global control information for multicellular development. CENEs are located in the the vast noncoding areas of genomes.  CENEs are not gene regulatory networks (GRNs) \cite{Carroll2008, Carroll2005, Davidson2006, Davidson2002, Werner2011a}.   Instead, CENEs subsume and control the gene regulatory networks (GRNs) that control the activation of genes. Thereby, CENEs control and organize cell actions such as division, movement and cell communication.  CENEs can be linked together to form larger more complex CENEs. The complete set of all CENEs in a genome is called the CENOME. The CENOME is the global control network that directs the development of Multi-Cellular Organisms (MCOs). 

\subsection{The Interpretive-Executive System IES and bilateral symmetry}
The cell has an {\em interpretive-executive system} (IES) that interprets the control information in the genome and executes its directives \cite{Werner2011a} to control cell actions in the dynamic system of interacting cells in a developing organism.  Part of the IES is the implicit coordinate system in the cell associated with its orientation and handedness.  Using cell orientation and its epigenetic interpretation by the IES, the theory of bilateral symmetry explains how bilaterally symmetric organism develop from a single cell \cite{Werner2012a}.   This theory of bilateral symmetry explains the pseudo-symmetric development of gynandromorphs.  Combining the developmental control theory of CENEs with the theory of bilateral symmetry,  explains all the varieties of existing, possible and yet to be engineered or discovered gynandromorphs. 

\section{Basic gynandromorphs}
\label{BasicGynanders}

\begin{figure}[H]
\centering
\subfloat[Part 1][Bilateral FM-FM-FM]{\includegraphics[width=\oriPicSize]{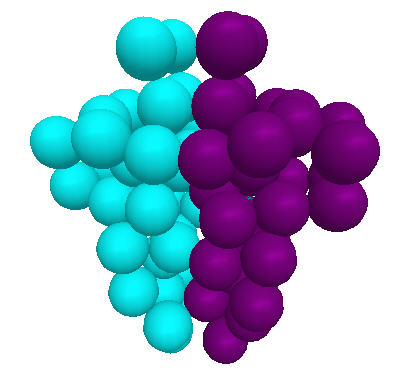}\label{fig:gynFbilatMCChrB}}
\subfloat[Part 2][Polar FF-FF-MM]{\includegraphics[width=\oriPicSize]{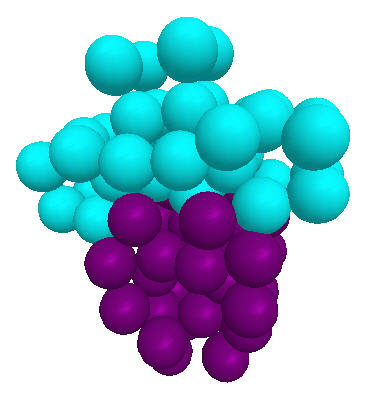} \label{fig:PolarFMb}}
\subfloat[Part 3][Oblique FM-FM-MF]{\includegraphics[width=\oriPicSize]{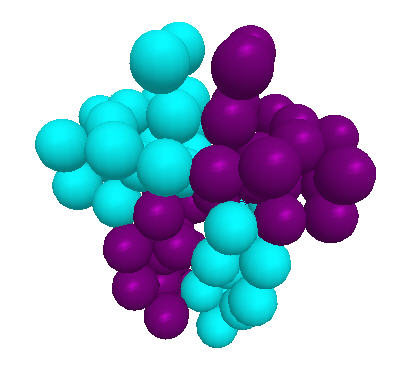} \label{fig:TrueObliqueCChrB}}
\subfloat[Part 4][Spiral MF-FM-FM]{\includegraphics[width=\oriPicSize]{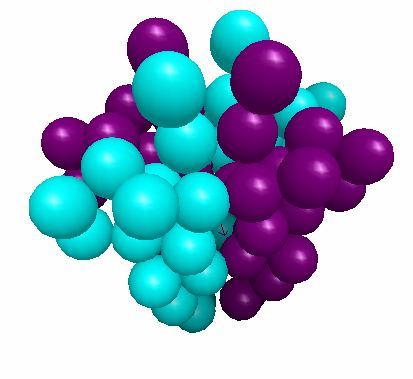}\label{fig:Spiral}} 
\caption{{\bf Basic gynandromorphs.} \it \small 
The first three figures illustrate the basic gynandromorph morphologies (bilateral, polar, oblique) found in mosquitos and other organisms. The last is a spiral gynandromorph that is a theoretical possibility. The organism has three main sections: Head, midsection and tail.  {\bf Notation:} F = Female, M = Male.  Going from anterior to posterior in Fig.\ref{fig:Spiral},  MF-FM-FM means: The bilateral head MF is right side Male and left side Female. The midsection FM is right side F and left side M. The posterior FM is right side F and left side M.  Let V = Ventral, D = Dorsal. When V and D are different as in the Spiral gynandromorph (Fig.\ref{fig:Spiral}) then it is more fully described by the pair (V: MF-FM-FM, D: FM-MF-MF)}
\label{fig:Gynander4Basic}
\end{figure}

Gynandromorphs  found in the wild come in at least three forms:  Bilateral\footnote{The term ``bilateral'' is used in several senses.  In its usual sense all the gynandromorphs (bilateral, polar and oblique) that we are modeling are bilateral organisms in the sense that they have developed bilaterally from a single cell.  In the second sense of bilateral, when used to classify gynandromorphs, it means that the male and female parts of the organism are separated into the two bilateral halves of the bilateral organism. Interestingly, there is a third sense of bilateral, that distinguishes bilateral development from the bilateral structure of an organism. While the gynandromorphs develop bilaterally from a single cell, phenotypically they are not necessarily bilateral in that the two sides of the bilaterally developing and developed organism need not be mirror images. The male parts in one bilateral half will have male characteristics different from the corresponding female structure in the other bilateral half.  However, the cell orientations of the two bilateral halves will mirror each other as described in \cite{Werner2012a}. }, polar, oblique \cite{Hall1988}.  
Using the computational theory of developmental control networks one can model and, starting from a single cell, simulate the embryonic development all the basic gynandromorphs found in nature (see \autoref{fig:Gynander4Basic}) as well as the all other possible forms that may exist (see \autoref{GynanderCombos}).  

\autoref{fig:Gynander4Basic}, illustrates the three basic forms of gynandromorphs: Bilateral gynandromorphs (bilateral body halves of opposite sex), polar gynandromorphs (anterior-posterior of opposite sex), and oblique gynandromorphs (opposite sex body sections cross the bilateral plane).  The spiral gynandromorph (when the dorsal D and ventral V sections of the organism are inter folded)  in Fig. \ref{fig:Spiral} is a theoretical possibility I have added the category of spiral gynandromorphs .  . 

At the same time gynandromorphs exhibit a pseudo symmetry in that they are bilaterally split down the middle with opposite handedness (see Fig. \ref{fig:GynaderArrows}).  The simulations of multicellular development of such organisms was done with software based on the theoretical framework. In the case of gynandromorphs the two bilateral founder cells have opposite orientations but now their control states activate distinct developmental control networks (CENEs).  Development then proceeds as if each side developed as part of a normal bilateral organism.  

\section{Bilateral gynandromorphs}
\begin{figure}[H]
\centering
\subfloat[Part 1][Normal female]{\includegraphics[width=\oriPicSize]{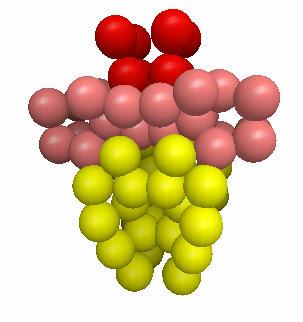} \label{fig:FemaleBase}}
\subfloat[Part 2][Gynandromorph]{\includegraphics[width=\oriPicSize]{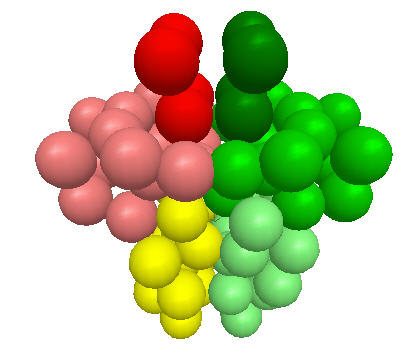} \label{fig:BilateralGynader}}
\subfloat[Part 3][Chromosome view]{\includegraphics[width=\oriPicSize]{gyn72-BilateralCChr.jpg} \label{fig:BilateralGynaderCChr}}
\subfloat[Part 4][Normal male]{\includegraphics[width=\oriPicSize]{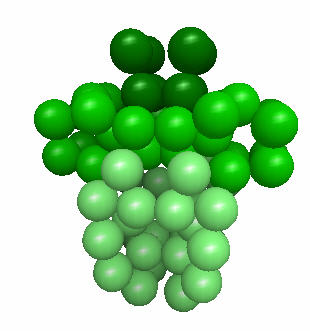} \label{fig:MaleBase}}
\caption{{\bf A bilateral gynandromorph.} \it \small 
The left figure Fig.\ref{fig:FemaleBase} is a normal female with a red head, orange midsection and yellow posterior.  The rightmost figure Fig.\ref{fig:MaleBase} shows the normal male, which in this example is morphologically identical to the female, but has different cell types which are indicated by shades of green.  The gynandromorph in Fig.\ref{fig:BilateralGynader} combines the male and female where the growth of each bilateral half is based, in part, on a different developmental control network. In this case, for the sake of clarity, the two halves are morphological mirrors of one another, just the cell differentiation states are different (as indicated by the different colors in Fig.\ref{fig:BilateralGynader}). The {\bf chromosome view} in figure Fig.\ref{fig:BilateralGynaderCChr}, of stained purple and aqua marine cells, shows which parental allelic genome, paternal or maternal, is active in each cell.  One half is female and the other male.  In principle, as long as the resulting embryo is viable, a gynandromorph-like organism could be a hybrid consisting of halves of two different organisms with distinct genomes.   }
\label{fig:Gynander3Views}
\end{figure}  

\subsection{Cell orientation in bilateral gynandromorphs}
Most gynandromorphs exhibit a pseudo bilateral symmetry in that each lateral half of the organism is the bilateral symmetric half of the normal organism.  Just as in the development of normal bilateral multicellular organisms, the cell orientation of gynandromorphs allows the consistent development of these pseudo bilateral halves of the gynandromorphs \cite{Werner2012a}.  

\begin{figure}[H]
\centering
\subfloat[Part 2][Cell differentiation]{\includegraphics[scale=0.3]{gynFbilatM.jpg} \label{fig:BilateralGynader2}}
\subfloat[Part 1][Cell orientation]{\includegraphics[scale=0.3]{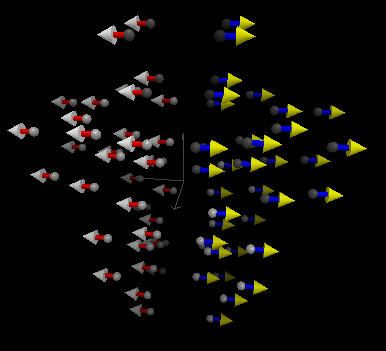} \label{fig:GynaderArrows}}
\subfloat[Part 2][Chromosome view]{\includegraphics[scale=0.3]{gyn72-BilateralCChr.jpg} \label{fig:BilateralGynaderCChr2}}
\caption{{\bf Three views of a bilateral gynandromorph.} \it \small 
There is a {\bf Cell Differentiation View} that shows the cell differentiation state by color, an {\bf Arrow View} that shows the cell orientations, and a {\bf Chromosome View} that shows which parental allelic genome, paternal or maternal, is active in each cell. {\bf Color codes:} In the Cell Differentiation View: Female cells F are red anterior, orange midsection, yellow posterior.  Male cells M are shades of green. In the Chromosome View: Female cells are aqua marine, Male cells are purple.  
}
\label{fig:Gynander3Views}
\end{figure}  

The growth of each bilateral half in \autoref{fig:Gynander3Views} is based, in part, on a different developmental control subnetwork. In these multicellular organisms the two halves are morphological mirrors of one another, just the cell differentiation states are different (as indicated by the different colors in the left Fig.\ref{fig:BilateralGynader2}).  One half is female and the other male.  But, in principle, a gynandromorph-like organism could consist of halves of two different organisms with distinct genomes as long as the resulting embryo is viable. The middle Fig.\ref{fig:GynaderArrows} shows the orientation states of the cells. Note the opposite orientation in the two body halves.  The right Fig.\ref{fig:BilateralGynaderCChr2}, of stained purple and aqua marine cells, shows which parental allelic genome, paternal or maternal, is active in each cell.

\subsubsection{Cell orientation in organ asymmetry}
This phenomenon of autonomous development of the two body halves also relates to asymmetric development. As in symmetric development, in asymmetric development the developing cells also have a handedness and orientation that is inherited epigenetically. This means that the switch in orientation of the asymmetric body part, e.g., left side to right side, is the result of a switch in orientation and handedness of the asymmetric founder, progenitor cell. This explains the ease and consistency of the switch of an organ or limb to its mirror, because the very same developmental control network (CENE) is being used to control the development of the mirror organ or limb.  All that needs to be changed is the orientation of the founder cell. 

Thus, early changes in cellular orientation have major developmental effects.  Hence, mutations in that lead to cell orientation switches can have vast evolutionary consequences \cite{Werner2012a}. 

\subsection{Avian bilateral gynandromorphs}
\label{sec:AvianGynanders}
It used be thought that sexual dimorphism, the difference in morphology of male and female animals, was due to hormones.  In a fundamental and important discovery it was shown that developmental differences in a chicken (gallus gallus domesticus) gynandromorph (Rooster-Hen in \autoref{fig:RoosterHen}) are cell based and not hormone based \cite{Zhao2010, Clinton2012}.  The cells on the different sides of the Rooster-Hen are of the opposite sex, the rooster half being male and the hen half being female.  In humans females have the homomorphic XX sex chromosomes and the males have the heteromorphic XY sex chromosomes. 

Unlike humans, chicken sex chromosomes have the opposite heteromorphic sorting. The cells of female chickens have the heteromorphic ZW sex chromosomes while the males have  homomorphic sex chromosomes ZZ.  Almost all of cells of the male half of the Rooster-Hen were male cells having the homomorphic ZZ sex chromosomes of roosters. The cells on the female half of the Rooster-Hen were female containing the heteromorphic WZ chromosomes of hens.  The same hormone has the opposite effect depending on the sex of the cells.  For example it induces male progenitor cells to become testes and female progenitor cells to become the female reproductive system \cite{Zhao2010, Clinton2012}.  
 Hence, a hormone which would be distributed evenly and effect both sides of the organism could not be responsible for the different phenotypes observed in the Rooster-Hen.  As a consequence the whole view of sexual dimorphism (differences in the phenotype of males and females) and how, at least avian, gynandromorphs originate has changed.  

\begin{figure}[H]
\centering\subfloat[Part 1][Rooster-Hen]{\includegraphics[scale=0.3]{RoosterHenClinton2.jpg} \label{fig:RoosterHen}} \hspace{1cm}
\subfloat[Part 2][Rooster-Hen Wattles]{\includegraphics[scale=0.65]{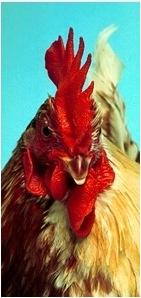} \label{fig:RoosterHenWattles}}
\caption{{\bf  The Rooster-Hen gynandromorph.} \it \small 
Fig.\ref{fig:RoosterHen} Rooster-Hen, right side female (brown), left side male. Picture provided by Clinton. A version appears in \cite{Zhao2010}.
Fig.\ref{fig:RoosterHenWattles} The same Rooster-Hen, showing the difference in wattles.  The right side is female (brown) with small wattles, left side is male with large wattles. 
 }
\label{fig:GynanderPics}
\end{figure}  

\subsubsection{Hormone information limits}
While it should have been obvious, even without this result, that a hormone, which is a relatively simple molecule, cannot contain the complex control information necessary for the development of the different complex morphologies that distinguish males and females.  The result of Clinton confirms the general theory that development is based on developmental control networks (CENEs) \cite{Werner2011a}.  And, it confirms the network theory of gynandromorphs which states that, as is the case in the Rooster-Hen, the two dimorphic body halves are controlled in part by different developmental control networks.  

\subsubsection{Prediction}
Since the Rooster-Hen preserves autonomous bilateral development, its cells are not only of the opposite sex and controlled by opposite sex developmental control networks, but the theory of bilateral symmetry predicts that they also are of opposite orientation and handedness \cite{Werner2012a}. Either early in development some process led to the generation of two founder cells of opposite orientation with mirror handedness. In the case of the Rooster-Hen gynandromorph  these founder cells were also of opposite sex. 

\section{Mosquito gynandromorphs}
Mosquitoes exhibit the basic palette of gynandromorphic types described in \autoref{BasicGynanders}, namely, bilateral, polar and oblique \cite{Hall1988}.  In this section we model the mixed polar-oblique form. 

\begin{figure}[H]
\centering
\subfloat[Part 1][Polar-Oblique Mosquito F]{\includegraphics[scale=0.4]{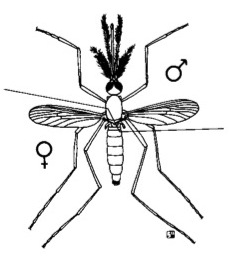} \label{fig:POMosquitoF}}
\subfloat[Part 2][Polar-Oblique Female]{\includegraphics[scale=0.4]{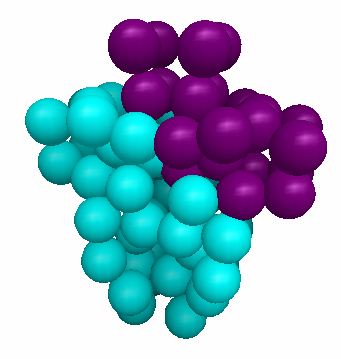} 
\label{fig:POMosquitoFMCO}} \hspace{0.6cm}
\subfloat[Part 3][MetaNet ]{\includegraphics[scale=0.2]{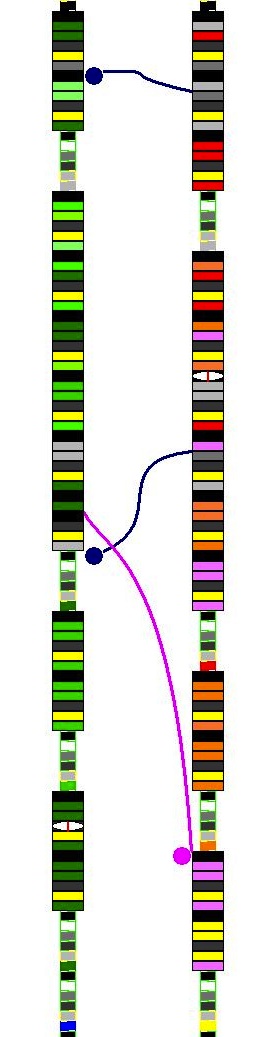} 
\label{fig:POMosquitoFMetaNet}} \hspace{0.6cm}
\subfloat[Part 4][Full Network ]{\includegraphics[scale=0.2]{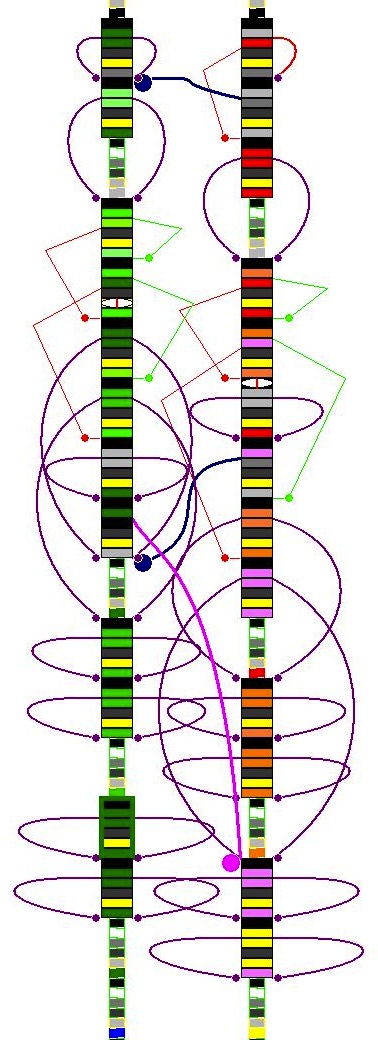} 
\label{fig:POMosquitoFNet}}\\
\subfloat[Part 5][Polar-Oblique Mosquito]{\includegraphics[scale=0.4]{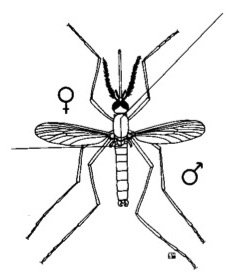} 
\label{fig:POMosquitoM}}
\subfloat[Part 6][Polar-Oblique Male]{\includegraphics[scale=0.35]{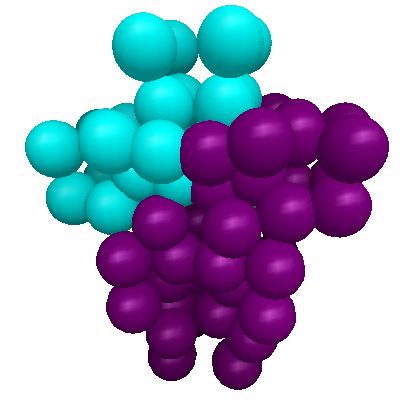} 
\label{fig:POMosquitoMMCO}}\hspace{0.6cm}
\subfloat[Part 7][MetaNet]{\includegraphics[scale=0.2]{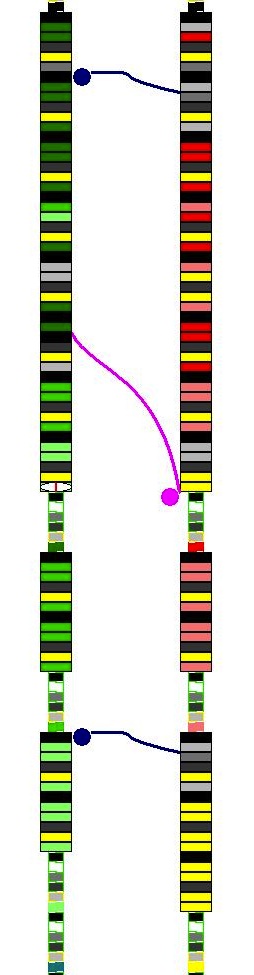} 
\label{fig:POMosquitoMMetaNet}} \hspace{0.6cm}
\subfloat[Part 8][Full Network]{\includegraphics[scale=0.2]{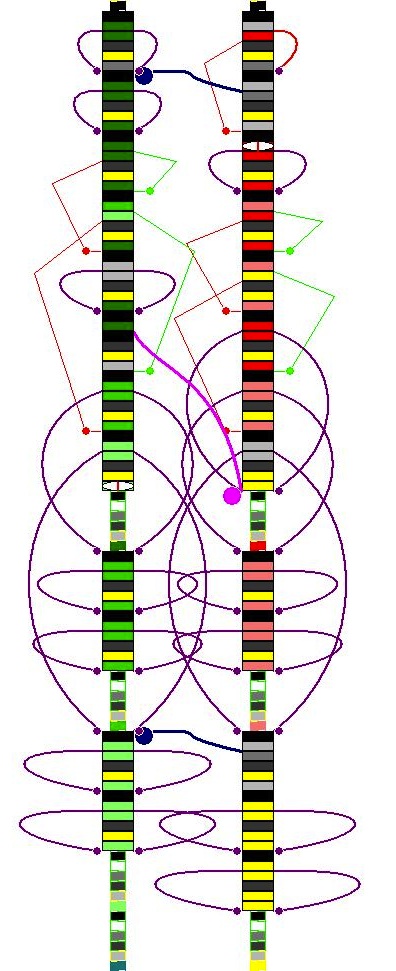} 
\label{fig:POMosquitoMNet}}
\caption{{\bf Polar-oblique mosquito gynandromorphs.} \it \small
Fig.\ref{fig:POMosquitoF} Polar oblique mosquito drawing from a real gynandromorph adapted from \cite{Hall1988}.  The head, right wing and midsection is male.  The left wing and midsection is female. The tail is female.  
Fig.\ref{fig:POMosquitoFMCO} A virtual polar oblique MCO with the same topology. 
Fig.\ref{fig:POMosquitoFMetaNet} The meta-network that generates the MCO. 
Fig.\ref{fig:POMosquitoFNet} The full network for the MCO. 
Fig.\ref{fig:POMosquitoM} Polar oblique mosquito drawing of a real gynandromorph adapted from \cite{Hall1988}.  The head, left wing and midsection is female.  The left wing and midsection is male. The tail is male.  
Fig.\ref{fig:POMosquitoMMCO} A virtual polar oblique MCO with the same topology. 
Fig.\ref{fig:POMosquitoMMetaNet} The meta-network that generates the MCO. 
Fig.\ref{fig:POMosquitoMNet} The full network for the MCO. Each parental genome is shown as one network with meta-links connecting the two allelic-genomes. 
 }
\label{fig:MosquitoPics}
\end{figure}  

In addition to sexually mixed morphology, gynandromorphs can have conflicting sex based behavior \cite{Maeno2007}. For example, polar gynandromorph mosquitos that have an anterior female half and a posterior male half.  The male abdomen is much smaller than the female's.  Such a polar gynandromorph mosquito has a female brain and will behave like a female sucking blood to nourish its nonexistent eggs. It will suck blood until its  small and inadequate male abdomen bursts \cite{Shetty1975}. 

%\pagebreak
\section{Gynandromorph combinatorics}
\label{GynanderCombos}

Beyond the basic gynandromorphs \autoref{BasicGynanders} there are many possible forms. 

\begin{figure}[H]
\centering
\subfloat[Part 1][Mid half female \\MM-FF-MM-MM]{\includegraphics[width=\oriPicSize]{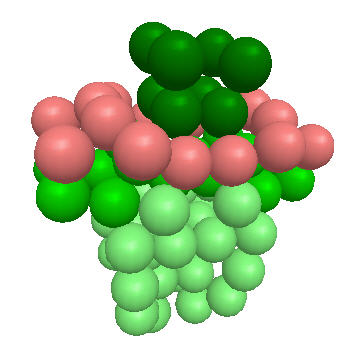} \label{fig:PolarMmidFM}}
\subfloat[Part 2][Mid male]{\includegraphics[width=\oriPicSize]{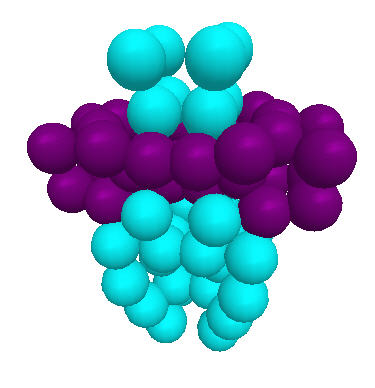} \label{fig:PolarMidM}}
\subfloat[Part 3][Symmetric male heart]{\includegraphics[width=\oriPicSize]{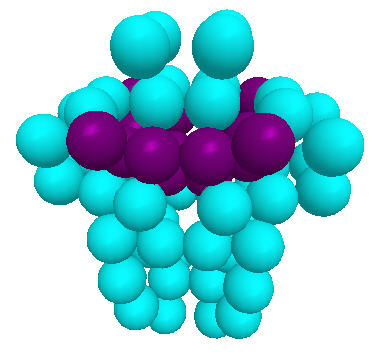} \label{fig:SymMaleHeart}}
\subfloat[Part 4][Spiral-oblique \\FM-FM-MF]{\includegraphics[width=\oriPicSize]{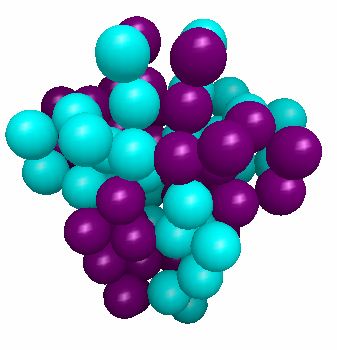} \label{fig:SpiralOblique}}\\
\subfloat[Part 5][Top quarter female]{\includegraphics[width=\oriPicSize]{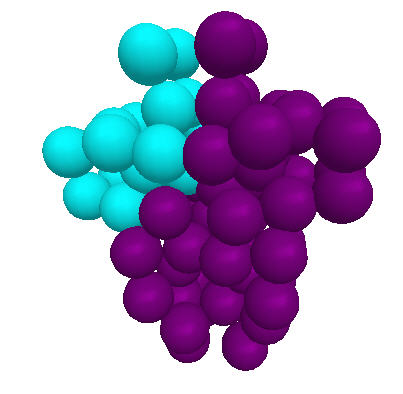} \label{fig:QuarterTopFemale}}
\subfloat[Part 6][Top quarter male]{\includegraphics[width=\oriPicSize]{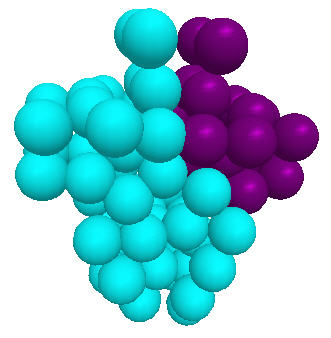} \label{fig:QuarterTopMale}}
\subfloat[Part 7][Quarter female]{\includegraphics[width=\oriPicSize]{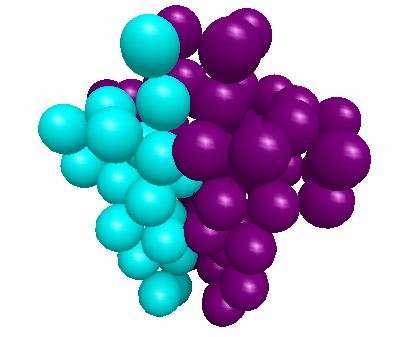} \label{fig:QuarterFemale}}
\subfloat[Part 8][Quarter male]{\includegraphics[width=\oriPicSize]{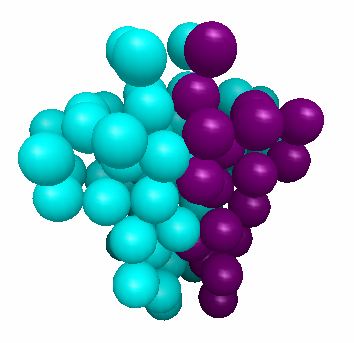} \label{fig:QuarterMale}}
\caption{{\bf Gynandromorph combinations.} \it \small  
Computationally modeled bilateral gynandromorphs each grown from a single cell (zygote) containing a set of male and female chromosomes.  The gynandromorphs are only two layers deep (dorsal and ventral) and have three anterior to posterior segments (head, midsection, tail). Illustrated some of the many possible combinations of gynandromorphs.  
Fig.\ref{fig:PolarMmidFM} shows a dominantly male gynandromorph with half of its midsection female. 
Fig.\ref{fig:PolarMidM} is a dominantly female gynandromorph with the whole midsection male. 
Fig.\ref{fig:SymMaleHeart} depicts a symmetric male ``heart'' in an otherwise female body.  
Fig.\ref{fig:SpiralOblique} is a spiral-oblique gynandromorph where male and female sections interlock between ventral and dorsal sides.  
Fig.\ref{fig:QuarterTopFemale} depicts a gynandromorph that has bilateral male female anterior and midsections with a polar male posterior. Viewed differently the top quarter is female and the rest male. 
In Fig.\ref{fig:QuarterTopMale} the top quarter is male. 
Fig.\ref{fig:QuarterFemale} shows a female gynandromorph where the whole front side is female in a male body. 
While Fig.\ref{fig:QuarterMale} shows the opposite. }
\label{fig:GynanderCombinations}
\end{figure} 

\autoref{fig:Gynander4Basic} and  \autoref{fig:GynanderCombinations} illustrate different forms of an idealized, very simple 3-segmented multicellular gynandromorph consisting of only 72 cells. The interactive, transactivation of male and female chromosomes lead to the phenotypes many of which are observed in nature. 

The gynandromorphs in \autoref{fig:Gynander4Basic} and  \autoref{fig:GynanderCombinations} are all bilateral multicellular organisms developed from a single cell.  They attain their different forms because the male (M) and female (F) chromosomes are differentially activated during development.  

\subsection{The possible developmental outcomes of gynandromorphs}  
It turns out the possible developments of gynandromorphs satisfy a combinatory logic.  Given each anterior to posterior section has two bilateral halves whose sex can vary independently,  each bilateral section can be in four possible states (MM, MF, FM, FF). Hence, if you just have two, anterior and posterior, body sections then if you start with MM you can get MM-MM (normal), MM-MF (polar-bilateral), MM-FM (polar-bilateral), MM-FF (polar).  If you start with MF you can have MF-MM (bilateral-polar), MF-MF (bilateral), MF-FM (oblique), MF-FF (bilateral-polar),  etc.  However, most insects like the mosquito have 3 main sections or segments, an anterior head segment, a midsection and a posterior, tail section.  Each has possibly different ventral (front) and dorsal (back) sides. Furthermore, each segment has distinct structure and possible further subsegments. 

In \autoref{fig:GynanderCombinations}, given 3 segments with 4 combinations (MM, MF, FM, FF) each, we have $4^{3} = 64$ possible gynandromorphs.  However, since we have two layers of cells, there is also a dorsal section D for each ventral section V.  When these are different as in Fig.\ref{fig:Spiral} =  (V: MF-FM-FM, D: FM-MF-MF), then, since both the ventral and dorsal sides of each section can vary independently, each section has $4 \times 4 = 16$ combinations. Hence, with 3 sections, we actually get $4^{6} = 16\times 16\times 16 = 64\times 64 = 4096$ possible gynandromorphs.  If we divide the midsection into two subsections, such as in the polar gynandromorph (V: MM-FF-MM-MM, D: MM-FF-MM-MM) in Fig.\ref{fig:PolarMmidFM}, we get $16$ more independent anterior-posterior, ventral-dorsal combinations to give a total of $4^{8} = 16 \times 4096 = 65,536$ combinations of gynandromorphs. Many of these transformations would be rather subtle and easily missed even when explicitly looking for gynandromorphic individuals. Others are very bizarre.  For example, one of these combinations is: (V: MM-FM-MF-FF, D: FM-FF-MF-MM).  

The organisms in \autoref{fig:GynanderCombinations} are part of an even larger space of possible gynandromorphs since the sections can also vary laterally within a bilateral half as in \autoref{fig:SymMaleHeart}   
giving at least  $4^{10} = 16 \times 65,536 = 1,048,576$ combinations of gynandromorphs.

\subsection{Open questions}
Given all these possible combinations each generating a different gynandromorph, why do we see only the major gynandromorphs?  Are we missing the rest because most changes are subtle?  Or is there a deeper organizational logic in the CENEs (developmental control networks)  underlying the development of the sections of insects?  

To what extent are the differences in sexual dimorphism determined by the sex chromosomes?  Are some of the CENEs that generate the sex based morphology in CENEs located on autosomes?

\section{Internetwork links between parental developmental network alleles leads to gynandromorphs}

The root cause of gynandromorphism are internetwork links between allelic developmental control networks responsible for sex differences lying on allelic chromosomes.  Each parent contributes a distinct set of allelic chromosomes that contain an allelic but different developmental network for an organism.  The mixture of these networks leads to the ontogeny of the organism.  When subnetworks responsible for the sexual dimorphism of a species are interlinked in the same developing individual organism, then that organism can exhibit aspects of both sexes.  The linkage causes a jump from a developmental parental network of one sex to the opposite sex developmental parental network.   This link leads from one network responsible for the morphology of one sex to the activation of the opposite sex morphology and function.  The more linkages there are between opposite sex developmental networks the more variable the gynandromorph phenotype.  

\begin{figure}[H]
\centering \subfloat[Part 3][Midsection male network]{\includegraphics[scale=\PicSzThree]{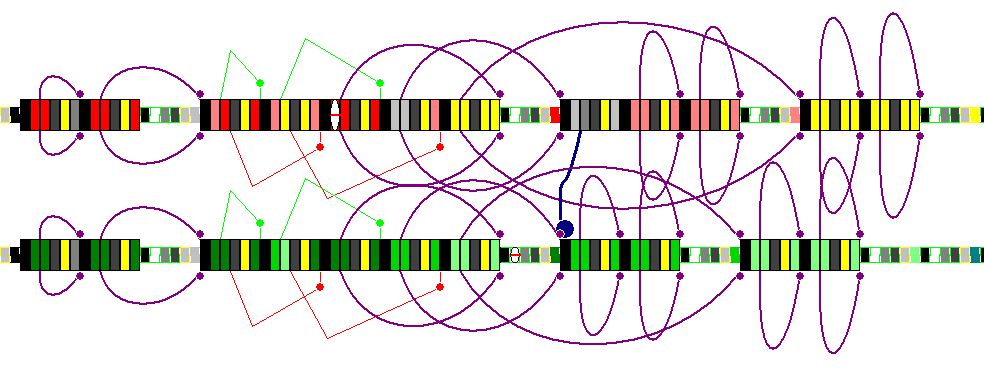} \label{fig:MidMaleMetaNet}} 
\subfloat[Part 4][Midsection male]{\includegraphics[scale=\PicSzGyn]{gyn72-MidMaleCCr.jpg} \label{fig:MidMale}} \\
\subfloat[Part 3][Midsection male meta-network signature]{\includegraphics[scale=\PicSzThree]{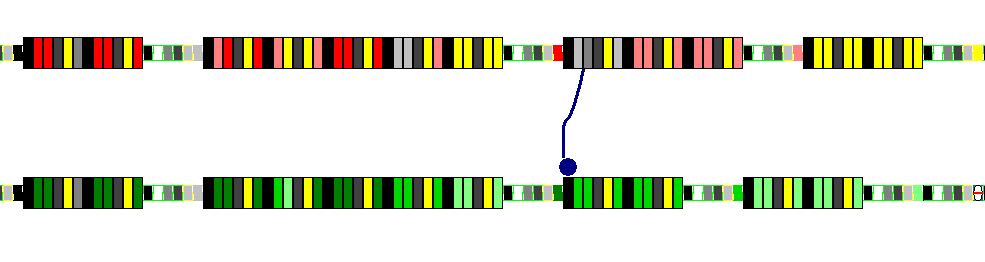} \label{fig:MidMaleMetaNet2}} 
\subfloat[Part 4][Midsection male]{\includegraphics[scale=\PicSzGyn]{gyn72-MidMaleCCr.jpg} \label{fig:MidMale2}} 
\caption{{\bf Two views of networks generating a gynandromorph with a male midsection.} \it \small 
The top network is the full view of the network including both the normal and meta-links while the bottom network shows only the meta-links between the two allelic parental genomes.  }
\label{fig:PolarNormalMetaNets}
\end{figure}  

%\pagebreak
\subsection{Meta-network interaction protocols between parental CENEs}
Each gynandromorph has a unique signature of meta-network links between parental developmental control networks. 

\begin{figure}[H] 
\centering 
\subfloat[Part 1][Bilateral gynandromorph meta-network]{\includegraphics[scale=\PicSzThree]{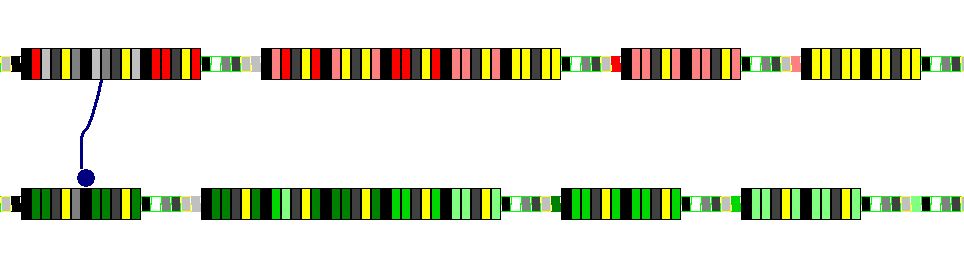} \label{fig:BilateralMetaNet}} 
\subfloat[Part 1][Bilateral gynandromorph]{\includegraphics[scale=\PicSzGyn]{gyn72-BilateralCChr.jpg}\label{fig:gynFbilatMCChrB2}}\\\subfloat[Part 2][Polar gynandromorph meta-network]{\includegraphics[scale=\PicSzThree]{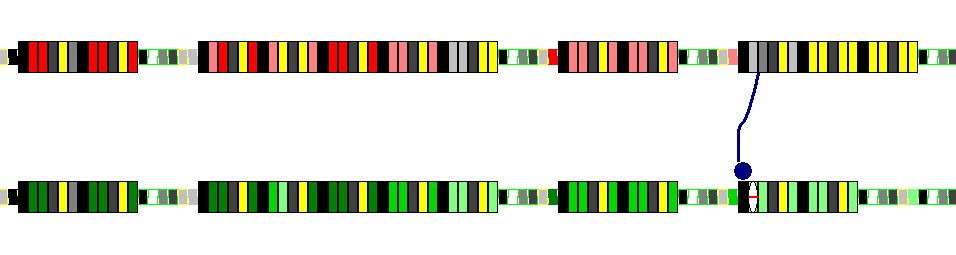} \label{fig:PolarMetaNet}} 
\subfloat[Part 3][Polar gynandromorph]{\includegraphics[scale=\PicSzGyn]{gyn72-PolarCChr.jpg} \label{fig:PolarGynander4}} \\
\subfloat[Part 4][Oblique gynandromorph meta-network]{\includegraphics[scale=\PicSzThree]{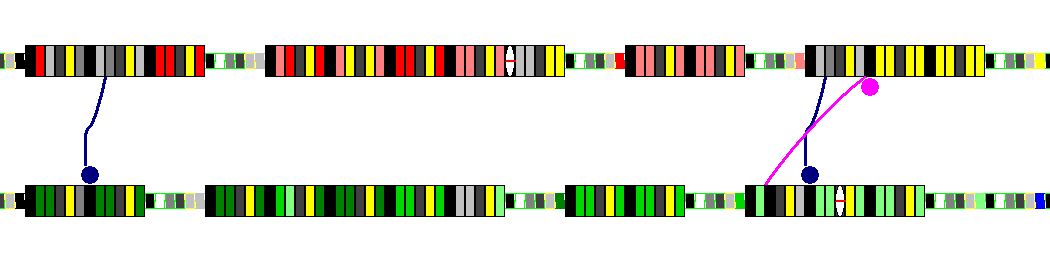} \label{fig:ObliqueMetaNet}}
\subfloat[Part 5][Oblique gynandromorph]{\includegraphics[scale=\PicSzGyn]{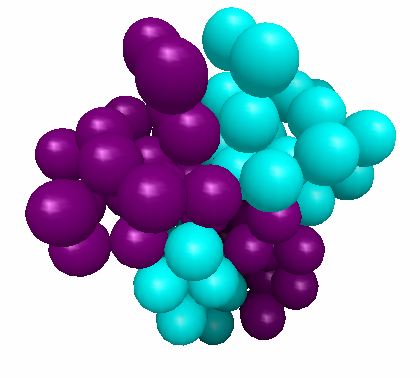} \label{fig:TrueObliqueCChr4}}\\
\subfloat[Part 6][Spiral gynandromorph meta-network]{\includegraphics[scale=\PicSzThree]{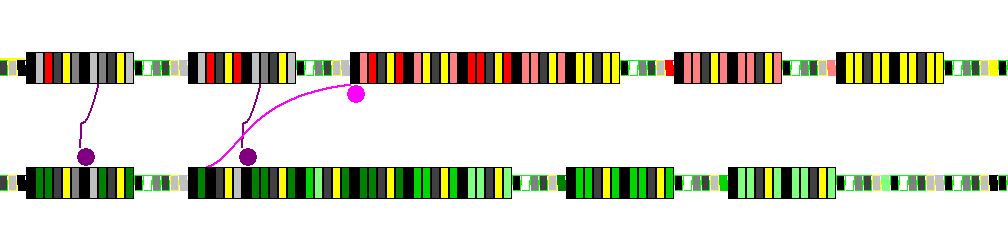} \label{fig:SpiralMetaNet2}} 
\subfloat[Part 7][Spiral gynandromorph]{\includegraphics[scale=\PicSzGyn]{gyn72-Spiral.jpg} \label{fig:Spiral4}}\\
\subfloat[Part 8][Spiral-oblique gynandromorph meta-network]{\includegraphics[scale=\PicSzThree]{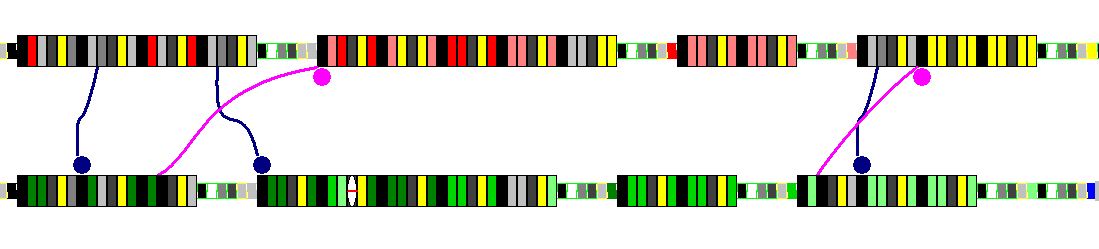} \label{fig:SpiralObliqueMetaNet}} 
\subfloat[Part 9][Spiral-oblique]{\includegraphics[scale=\PicSzGyn]{gyn72-SpiralOblique.jpg} \label{fig:SpiralOblique2}}
\caption{{\bf Signatures of meta-networks that generate basic gynandromorphs.} \it \small 
The main differences in these networks are the meta-control linkages between the primary developmental networks. Each genome has a unique signature of network links between its complementary parental haploid genomes. All the genomes are identical except for their distinct inter-haploid network linkages. }
\label{fig:ManyMetaNets}
\end{figure} 

Irrespective of the number of combinations, what is interesting about these combinations is that they indicate an underlying global organization in the genome and its cenome.  It leads me to suspect that the genome-cenome has a well organized architecture with sections of developmental control networks partitioned to be in correspondence with the major sections that vary in gynandromorphs. 
 
\subsection{Deterministic versus stochastic meta-networks}
The links between opposite sex developmental control networks may be deterministic or stochastic.  If they are deterministic then the same gynandromorphic phenotype will develop in all individuals whose cells are controlled by that network. If the links are stochastic and occur with low probability then we get what we see in the case of mosquitos were only a very small number of individuals appear to be gynandromorphs.  However, because many gynandromorphs may be subtle variations of the millions of possibilities that are not overtly obvious, there may exist many more gynandromorphs than have been recognized as such.  Hence, the chromosomal link architecture between parental CENEs may be more complex than indicated by the overt cases of insect gynandromorphs. 

\subsection{Bioengineering gynandromorphs}
Given the existence of stochastic linkages between the developmental control networks of sex chromosomes, one should be able to engineer deterministic inter-sex developmental network links that replace stochastic links in the allelic developmental control networks of an insect like the mosquito. This would create a deterministic gynandromorphic network which if viable would lead to a specific gynandromorphic phenotype. 

\subsection{Genetic constraints on the space of possible gynandromorphs} 
The standard view is that variations in chromosome partitioning via unequal chromosomal placement in daughter cells creates  gynandromorphs.  However, the work of Clinton has challenged this view (see \autoref{sec:AvianGynanders}).  Still on both accounts, for sexually dimorphic gynandromorphs both the male and female CENEs that control dimorphic differences in development must exist in the gynandromorph's genome. 
Chromosome loss or duplication can lead to inadvertent activation of the opposite sex chromosome. This is equivalent to a meta-network link being formed between the different maternal and paternal developmental networks. 

We have shown above that  chromosomal activation by cross chromosomal linking is an alternative way to generate gynandromorphs.  
The meta-network signatures that distinguish various types of gyn72 gynandromorphs assume a normal compliment of male and female CENEs, e.g., XY or ZW heteromorphic chromosomes.  The gynandromorphs result from switches of control between one CENE and the sex-opposite CENE.  This can be engineered into the CENEs by the creation a cross chromosomal link. It can result of a mutation that switches a link to point to the homologous area of the opposite sex chromosome.  Or, it can occur by different chromosomal placement in pairs of progenitor cells as the organism develops.  

Chromosomal placement can put limits on the future possible chromosomal divergence in daughter cells. For example, if XY produces an XX-cell and a YY-cell then we get one normal female cell and an abnormal male YY-cell. But neither of these cells can now generate a future mixed XY cell or cell of the opposite sex since the opposite sex complement is missing.  Hence, once such a chromosomal separation occurs, if sex dimorphism is determined by CENEs on the X and Y chromosomes exclusively without the help of CENEs on autosomes then we can no longer get further gynandromorphic differences.  Thus if development establishes an XX bilateral half that half can no longer generate a male Y based subsection.  Whereas an XY bilateral half still has the potential to generate an XX female subsection.  Therefore, chromosome placement limits the types of possible gynandromorphs.  

In the network perspective, chromosome placement limits the possible meta-networks to a proper subset of the set of all possible meta-networks between sex based chromosomes.  However, the early development of the embryo can influence chromosomal placement so that one may still get the basic gynandromorphs including bilateral, polar and oblique.  Hence, the set of possible gynandromorphs resulting from chromosomal placement is relative to the upstream developmental network (upstream sub-CENE).   

\section{Synsexhemimorphism:  Female and male hemimorphic organisms}
In a female that contains two X chromosomes, normally one of the X chromosomes is suppressed.  If not, we get abnormal development. The network theory of gynandromorphs predicts that there can exist females and males in various gynandromorph-like forms but where the bilateral, polar, or oblique versions are not governed by opposite sex chromosomes, but  by different same sex chromosomes and/or autosomes from different parents.  Similar can problems occur if the organism has two male Y chromosomes.  

\begin{figure}[H]
\centering 
\subfloat[Part 1][Bi-Female Phenotype]{\includegraphics[scale=\PicSzThree]{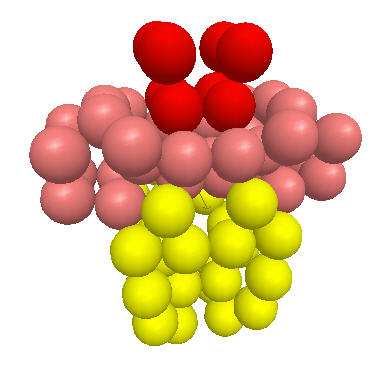} \label{fig:GynF}} 
\hspace{1.0cm}
\subfloat[Part 2][Bi-Female Chromosome view]{\includegraphics[scale=\PicSzThree]{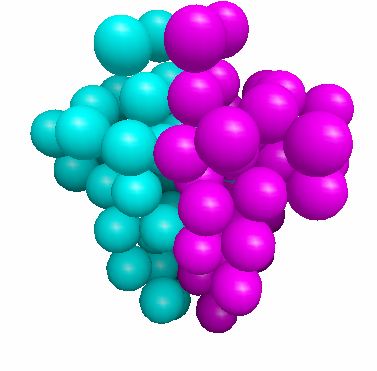} \label{fig:GynaderFFCChr}}
\caption{{\bf Two views of a bilateral bi-female organism.} \it \small 
This female looks normal with all three sections being female.  But actually, as the right chromosome view shows, each bilateral half of the embryo has developed from different developmental networks located on a different chromosomes.  Hence, there is the possibility of an organism developing that is female, but with two distinct female ancestors generating the two bilateral body halves.  Polar, oblique and spiral versions are, of course, also possible. Hence, a person might inherit the head  from their maternal grandmother, the midsection from the paternal grandmother and the lower half from their maternal grandmother.  }
\label{fig:BiFemale}
\end{figure}  

Thus, under the network theory of gynandromorphs there may exist organisms that are of the same sex (are not gynandromorphs) but that have non symmetrical or non-coherent body sections because their development was directed by distinct but same-sex developmental control networks.   

{\bf Definitions}:   {\em Synsexhemimorphs } are same sex (male or female) organisms with asymmetric development, of normally symmetric body parts/sections, resulting from different parental allelic, developmental control networks ( CENEs).  
{\em  Gynohemimorphs } are overtly female organisms with asymmetric body parts/sections resulting from different parental allelic, developmental control networks or CENEs. 
{\em  Androhemimorphs } are overtly male organisms with asymmetric body parts/sections resulting from different, allelic parental developmental control networks or CENEs.  

\subsection{Hemihyperplasia as synsexhemimorphism } 
In the case of human males because they have only one Y chromosome one might think that diandromorphs cannot exist.  However, homologous, allelic autosomes (and perhaps even the X chromosome)  will usually contain nonidentical developmental control networks.  Two homologous allelic CENEs on homologous autosomes may thus conflict in their phenotype and if asymmetrically activated this phenotypic disparity would become evident.  

Examples of synsexhemimorphs (androhemimorphs or gynohemimorphs) may be humans with hemihyperplasia  (hemihypertrophy) which is defined by asymmetric growth of normally symmetric body parts or sections including the cranium, face, trunk, limbs, and/or digits.  This may be a form of developmental network induced male or female hemimorphism.  Note, under this hypothesis hemihyperplasia need show neither chromosomal nor genetic abnormalities, because it is a result of meta-network cross talk between homologous, allelic autosomal developmental networks (CENEs). 

\section{Control conflicts and the origin of species}

This opens up a much more general and fundamentally important question:  How is the contribution of two different parents, their two different developmental control networks controlled?  What determines which of the two different parental networks acts when and where?  

Since these organisms are bilaterally symmetric the meta-control network that specifies which network is in charge at each point in space and time in development cannot be a random or stochastic process.  Otherwise, if it were stochastic then the two bilateral sides would be  morphologically dissimilar and not bilaterally symmetric.  So too, identical twins show that the protocol of interaction (a meta-network) between the two parental genomes containing divergent developmental networks, cannot be random.  

\subsection{Sexual cooperation via meta-network protocols between parental genomes}
Furthermore, the existence of gynandromorphs appears to indicate that control of which parental network is active at any given point in development cannot be a simple compromise between both networks that takes the average of two conflicting directives to a given cell at each point in time.  

If this line of reasoning is correct, then it follows that there must exist a meta-control network that interlinks the two parental developmental control networks.  This meta-control architecture implements a control protocol that coordinates the action of the two parental genomic contributions may be universal for a given species. Conflicts between such meta-control network linkages may be the cause of the very existence species.  

Hence, the existence and origin of species is primarily the result of divergent developmental control networks and specifically divergent meta-control networks that interlink primary developmental networks. Thus, the evolution of species is the result of developmental network transformations.  

\subsection{Evolution of species}
Since coordination conflicts would likely lead to unviable offspring, difference in the meta-control architecture, and hence, different meta-control linkages between parental networks, partition organisms into distinct classes, thereby creating different species.  Hence, the origin of species results from unresolvable control conflicts between the developmental networks of different, putative species.  A new species forms when the meta-control linkages between the parental networks of the new and the old species are no longer compatible resulting pathological development and/or unviable embryos. 

The difference between species is a matter of degree. The greater the difference between the developmental networks of two species coordinating the networks of two divergent species becomes difficult no matter what meta-control linkages exist. Thus, if all possible meta-linkages (other than the null link and trivial links) between two parental networks fail to produce a viable offspring, then the species are very distant.  Once, we decode the syntax and semantics of genomes \cite{Werner2003,Werner2005}, we will have a new way of tracing the origin and evolution of species based on the architecture of developmental control networks. 

\subsection{Reverse engineering species formation}
Just as it may be possible to engineer gynandromorphs, so it may be possible to engineer new species by modifying the meta-network interaction protocol or signature between parental CENEs.  Also given two different species it may be possible to reverse engineer the two species by transforming their meta-network signatures to be compatible.  This could make at least artificially induced fertilization between formerly distinct species possible and the resulting embryonic development viable.  Hence, formerly distinct species A and B would have their meta-network signatures modified to A* and B* so that their interactions A*xB* leads to a viable embryo. However, would A*xA* or B*xB* still be viable? 

Alternatively, it may be possible to reverse engineer an ancient ancestor species by transforming the meta-network signature to that of an ancestor. The case of atavisms shows that this is in principle possible. The feasibility depends on which ancient developmental control subnetworks are still there, hidden in the organism's genome.  By combining related specie's CENEs using meta-network signature transformations one may be able to reverse engineer the common ancestor CENE taking parts from each of the two or more species CENEs.  For example, could avian (bird) CENEs meta-linked to reptile CENEs by reverse species signature transformations allow the engineering of dinosaur like organisms? 

\section{Conclusion: How it all fits together}
 
Gynandromorphs develop the way they do because of an interdependent cooperation between parental developmental control networks (CENEs), gene regulatory networks (GRNs), epigenetic cell orientation, cell-cell interconnection physics, and cell communication.  CENEs  subsume GRNs to control cell action. The cell's interpretive-executive system (IES) interprets the control information in the genome (its cenome) relative to the cell's orientation and executes that information using the GRNs to activate genes to perform various actions such as cell division, cell signaling, cell movement, and cell differentiation.  Cell physics of intercellular connections also plays its role in the ultimate outcome.  Cell communication is involved throughout development for spatial and temporal error correction, cell-cell coordination and cooperation.  

Cells in the opposing symmetric body halves of bilateral organisms have opposite orientations and handedness. 
Handedness once established is network autonomous unless and until it is changed.  
Gynandromorphs show that the handedness of development of one half of the multicellular organism MCO1 is not dependent on a  particular developmental control network N1 because the other oppositely oriented half MCO2 of the gynandromorph can develop according to a different developmental control network N2.
Epigenetic cell polarity and not the genome is the primary cause of bilateral symmetry since cells maintain and inherit their orientation and handedness epigenetically. Cell signaling may be used to maintain polarity and orientation. 

Since the same initial cell type can lead to two different morphologies in the two bilateral halves of gynandromorphs, the genome and not the containing cell is the prime driver of multicellular morphology. 
Gynandromorphs confirm the relative independence of morphology and cell orientation. It shows the relative independence of the morphological control information contained in the genome and the orientational information contained in the cell. 
Developmental control networks and not hormones determine the morphology of each bilateral half of a gynandromorph.
Divergent subnetworks of the global developmental control network (the cenome) and not hormones determine the distinct morphology and function of sexually dimorphic organisms.  Unequal chromosomal partitioning between cells can cause gynandromorphy because such partitioning can lead to modified meta-network linkages between allelic parental networks (whether they are located on sex chromosomes or autosomes). 

The parental allelic developmental control networks must cooperate via meta-network linking protocols in order for coherent development to take place.  Species form when established meta-network links change or are replaced or supplemented by new internetwork links that make sexual network cooperation between putative parental genomes unlikely or lead to unviable or infertile offsprings.  Gynandromorphs led us to the insight that not genetic differences but rather developmental network divergences and incompatible meta-networks, which link the haploid developmental networks of the two parental sexes, result in the origin of species.  

\section{Materials and methods}
\label{Methods}

Cellnomica's Software Suite (http://cellnomica.com) was used to model and simulate gynandromorph multicellular development in space-time.  The gynandromorphs in all the figures were all tested using Cellnomica's Software Suite. Each of the concepts discussed was  modeled and simulated with artificial genomes that generated multicellular bilaterally symmetric gynandromorphs starting from a single cell.  Both stochastic and deterministic gynandromorph networks were developed and tested.  The illustrations of multi-cellular systems   are screenshots of cells that developed in virtual 4-dimensional space-time , modeled using the Cellnomica's software. 
The illustrations of developmental control networks are screenshots of networks modeled and run with Cellnomica's software. 

\nocite{*}

\addcontentsline{toc}{section}{References}% for showing references in table of contents 
\begin{multicols}{2}

\footnotesize 
\bibliographystyle{abbrv}
\bibliography{EWernerGynandromorphs}

\end{multicols}
\end{document}